\documentclass[12pt,preprint]{aastex}
\usepackage{epsf}

\newcommand{\be}{\begin{displaymath}}
\newcommand{\ee}{\end{displaymath}}
\newcommand{\bea}{\begin{eqnarray}}
\newcommand{\eea}{\end{eqnarray}}

\shortauthors{Pavel Denissenkov}
\shorttitle{A New Twist In the Evolution of Low-Mass Stars}

\begin{document}

\title{A NEW TWIST IN THE EVOLUTION OF LOW-MASS STARS}

\author{Pavel A. Denissenkov}
\affil{Department of Physics \& Astronomy, University of Victoria,
       P.O.~Box 3055, Victoria, B.C., V8W~3P6, Canada}
              \email{pavelden@uvic.ca}
 
\begin{abstract}
We show that the evolutionary track of a low-mass red giant
should make an extended zigzag on the Hertzsprung-Russel diagram just after the bump luminosity,
if fast internal rotation and enhanced extra mixing in the radiative zone
bring the temperature gradient close to the adiabatic one. This can explain
both the location and peculiar surface chemical composition of Li-rich K giants
studied by \cite{kea11}. We also discuss a striking resemblance between the photometric and composition peculiarities of
these stars and giant components of RS CVn binaries. We demonstrate that the observationally constrained values of
the temperature gradient in the Li-rich K giants agree with the required rate of
extra mixing only if the turbulence which is believed to be responsible for this extra mixing is highly anisotropic,
with its associated transport coefficients in the horizontal direction strongly dominating over those in the vertical direction.
\end{abstract} 


\section{Introduction}

During their first ascent along the red giant branch (RGB), low-mass stars
are known to experience extra mixing in their convectively stable
radiative zones, between the hydrogen burning shell (HBS) and the bottom of the convective envelope (BCE).
This occurs above the bump luminosity (\citealt{grea00}), after the HBS has crossed and erased 
a chemical composition discontinuity left behind by the BCE at the end of the first dredge-up (FDU).
As a result, the atmospheric $^{12}$C/$^{13}$C ratio and, in metal-poor stars, carbon abundance
resume their post-FDU declines with increasing luminosity. Extra mixing is needed
to explain the observed low carbon isotopic ratios in evolved stars with initial masses up to $2.2\,M_\odot$
(\citealt{chl10}). Another abundance anomaly that has also been associated with the start of RGB extra mixing
at the bump luminosity is the lithium enrichment that has been detected in a small fraction (1\,--2\,\%) of K giants
(\citealt{chb00}). To produce $^7$Li via the $^7$Be mechanism (\citealt{cf71}), extra mixing has to be 
nearly three orders of magnitude faster in these stars than in other low-mass red giants
above the bump luminosity (\citealt{sb99,dw00}). 

The problem of the origin of Li enrichment in K giants has recently been addressed again by \citet[hereafter KRL]{kea11}.
They observed 2000 K giants with determined Hipparcos parallaxes and found that 15 of them are Li-rich.
However, the most interesting new result is that many of these Li-rich K giants are located
well below the bump luminosity, close to the red horizontal-branch clump region of the Hertzsprung-Russell diagram (HRD),
where the low-mass stars should arrive much later, after
they will have experienced the He-core flash. The newly discovered Li-rich K giants also exhibit
very low $^{12}$C/$^{13}$C ratios, approaching the equilibrium value for the CN cycle in four of them. 
Analysing these data, KRL have proposed that the lithium enrichment in these K giants occurred during the He-core flash
rather than at the bump luminosity.

In this Letter, we present an alternative explanation; namely, we assume that these stars
have already reached the bump luminosity and, therefore, that RGB extra mixing is now operating in their radiative zones. However,
for some reason, presumably because of their very rapid internal rotation, this mixing is so efficient that
not only can it trigger the $^7$Be mechanism, but its driving force or associated heat transport can also modify
the radiative zone's thermal structure, resulting in a temperature gradient that is closer to the adiabatic one.
In this case, a bump-luminosity star in which such enhanced extra mixing has just started should 
make an extended zigzag on the HRD towards
much lower luminosities, comparable to those of red clump stars, before resuming its ascent along the RGB (\citealt{dea06,dp08b}).
This hypothesis is supported by a striking resemblance between the photometric and chemical 
composition peculiarities of the primary (red-giant) components of the RS CVn binaries and the Li-rich K-giants studied by KRL.

\section{Photometric Peculiarities Caused by Enhanced Extra Mixing}
\label{sec:photopec}

The RS CVn binaries are close stellar systems in which the primary components are low-mass red giants.
In many cases, the RS CVn giant components have synchronized their spin and orbital periods and, as a result of
their fast rotation caused by the tidal spinning up, they show signatures of chromospheric activity (\citealt{fea02,mea04,fh05}). 
\cite{mea04}, and \cite{dea06} have noted a remarkable photometric peculiarity intrinsic to the RS CVn giants 
and other K-giants in tidally locked binaries --- most of them reside on the lower RGB, below the bump luminosity.

\cite{dp08b} have shown that, after having reached the bump luminosity, a low-mass star will make a long excursion
towards much lower luminosities if the temperature gradient $\nabla\equiv (d\ln T/d\ln P)$
in its radiative zone increases and takes a value between the radiative and adiabatic temperature gradients,
$\nabla_{\rm rad}<\nabla <\nabla_{\rm ad}$. 
Such a deviation of $\nabla$ from $\nabla_{\rm rad}$ can be explained in two ways.
On the one hand, a rotational distortion of level surfaces in the radiative zone leads to an increase of 
$\nabla =\nabla_{\rm rad}$ to $\nabla\approx (1+2\varepsilon)\nabla_{\rm rad}$, where
$\varepsilon\approx (\Omega^2r^3/3GM_r) < \varepsilon_{\rm crit}$ is the ratio of the equatorial centrifugal acceleration to
gravity divided by 3, and $\varepsilon_{\rm crit} = 0.24$ is its critical value at which the Roche equipotential surface
breaks up at the equator (\citealt{dvb03b}). It is this effect that was proposed by \cite{dea06} to be responsible
for the photometric peculiarity of red giants in tidally locked binaries. On the other hand,
fast differential rotation and its related hydrodynamic or magnetohydrodynamic instabilities
can be driving mechanisms for extra mixing in stellar radiative zones (e.g., \citealt{z92,s99}).
Besides chemical elements, this extra mixing should also transport heat. The efficiency of the latter process
in the standard mixing length theory is estimated by the quantity $\Gamma\approx\gamma (D_{\rm mix}/K)$,
where $D_{\rm mix}$ is the rate (diffusion coefficient) of extra mixing, $K$ is the thermal diffusivity, and $\gamma$
is the factor determined by the geometry of the fluid elements. In this case, the temperature gradient
lies between $\nabla_{\rm rad}$ and $\nabla_{\rm ad}$, and its value is given by
$\nabla\approx (1-f)\nabla_{\rm rad} + f\nabla_{\rm ad}$, where $f = 6\Gamma^2/(1+\Gamma +6\Gamma^2)$ (\citealt{m95}). 

\begin{figure}
\epsffile [60 190 480 695] {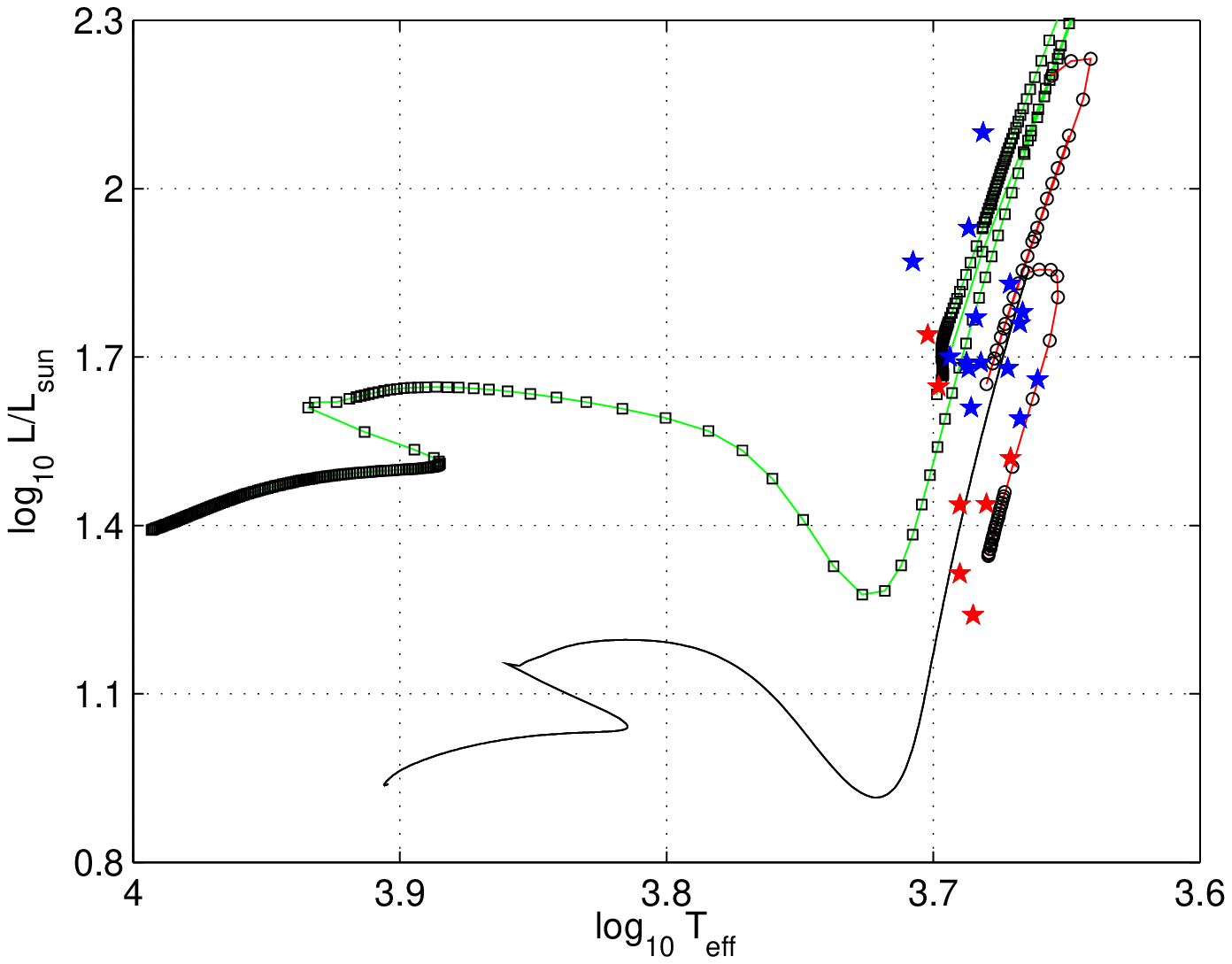}
\caption{The black and green curves are the standard evolutionary tracks of the $1.7\,M_\odot$ and $2.2\,M_\odot$ 
         solar-composition stars. The red curves show the bifurcations caused by convective heat transport with 
         the efficiencies $\Gamma = 1$ (the lower curve) and $\Gamma = 0.75$ (the upper curve) that starts in the 
         radiative zone above the bump luminosity. As a result, the stars make
         excursions towards lower luminosities before resuming their ascents along the RGB.
         The blue and red star symbols represent, in turn, the Li-rich K giants studied by KRL and the giant components of
         RS CVn binaries. The neighbouring open symbols are separated by one million years.
         }
\label{fig:f1}
\end{figure}

The giant components of the RS CVn binaries selected by us from the paper of \cite{mea04} for comparison and 
the Li-rich K giants discovered by KRL both have nearly solar metallicities and masses close to $2\,M_\odot$.
Therefore, we compare their positions on the HRD (the red and blue star symbols in Fig.~\ref{fig:f1}) with the evolution
of two stellar models calculated\footnote{All the stellar evolution computations for this paper have been done with 
the MESA code (\citealt{pea11}) that is freely available at {\tt http://mesa.sourceforge.net}.}
for the solar heavy-element and helium mass fractions, $Z=0.019$ and $Y=0.28$, 
and initial masses $1.7\,M_\odot$ and $2.2\,M_\odot$ (the black and green curves in the same figure). 
The neighbouring open symbols along the evolutionary tracks are separated by one million years. Their
concentrations on the green curve at low $T_{\rm eff}$ are increased at the bump luminosity, $\log_{10}(L/L_\odot)\approx 2.25$, and
at the red clump, $\log_{10}(L/L_\odot)\approx 1.7$. Before the star has evolved
beyond the bump luminosity, its radiative zone contains a strong gradient of the mean molecular weight,
$\nabla_\mu\equiv d\ln\mu/d\ln P > 0$, that is believed to prevent any extra mixing. The blue star symbols are all located below
the bump luminosity, in a region of the HRD that the standard evolutionary track (the green curve) of a low-mass red giant
with a chemically uniform radiative zone can reach only after the He-core flash. 
This explains why KRL have inferred that
both the production of Li and the reduction of the carbon isotopic ratio in their Li-rich K-giants
occurred during the He-core flash.

However, the $^7$Be mechanism of surface Li enrichment requires strongly enhanced extra mixing with a rate (diffusion coefficient)
increased by nearly three orders of magnitude compared to that of canonical extra mixing operating in   
the majority of RGB stars above the bump luminosity (\citealt{dvb03a}). We surmise that
such mixing or its associated fast rotation should modify $\nabla$ in the radiative zone
bringing its value closer to $\nabla_{\rm ad}$. The red curve originating 
from $\log_{10}(L/L_\odot)\approx 2.25$ in Fig.~\ref{fig:f1}
shows a bifurcation of the evolution of the $2.2\,M_\odot$ star after the bump luminosity caused by
our assumption that the start of enhanced extra mixing with $D_{\rm mix} = 2\times 10^{11}$\,cm$^2$\,s$^{-1}$
in this star has initiated a convective heat transport in its radiative zone with the efficiency
$\Gamma = 0.75$. It is seen that the star spends quite a long time at luminosities close to those of the Li-rich K-giants.
We have assumed that the enhanced extra mixing has the same maximum depth, $r_{\rm mix} = 0.05\,R_\odot$,
as the canonical one (\citealt{dp08a}). Before the HBS, advancing in mass outwards, has crossed 
the chemical composition discontinuity left behind by
the BCE at the end of the FDU, the mixing is allowed to operate only in the chemically uniform zone between the BCE and 
the current location of the discontinuity at $r>r_{\rm mix}$. 
Note that it is the increase of $\nabla$, not the effect of mixing, that forces the star to make the extended zigzag. 
Fig.~\ref{fig:f1} also shows that the region occupied by the majority of tidally locked giants in close binaries, 
including the RS CVn giant components
(red star symbols), is reached by the evolutionary U-turn of the $1.7\,M_\odot$ star (red curve) when $\Gamma = 1$.

\section{Chemical Composition Peculiarities Caused by Enhanced Extra Mixing}
\label{sec:chempec}

\begin{figure}
\epsffile [60 190 480 695] {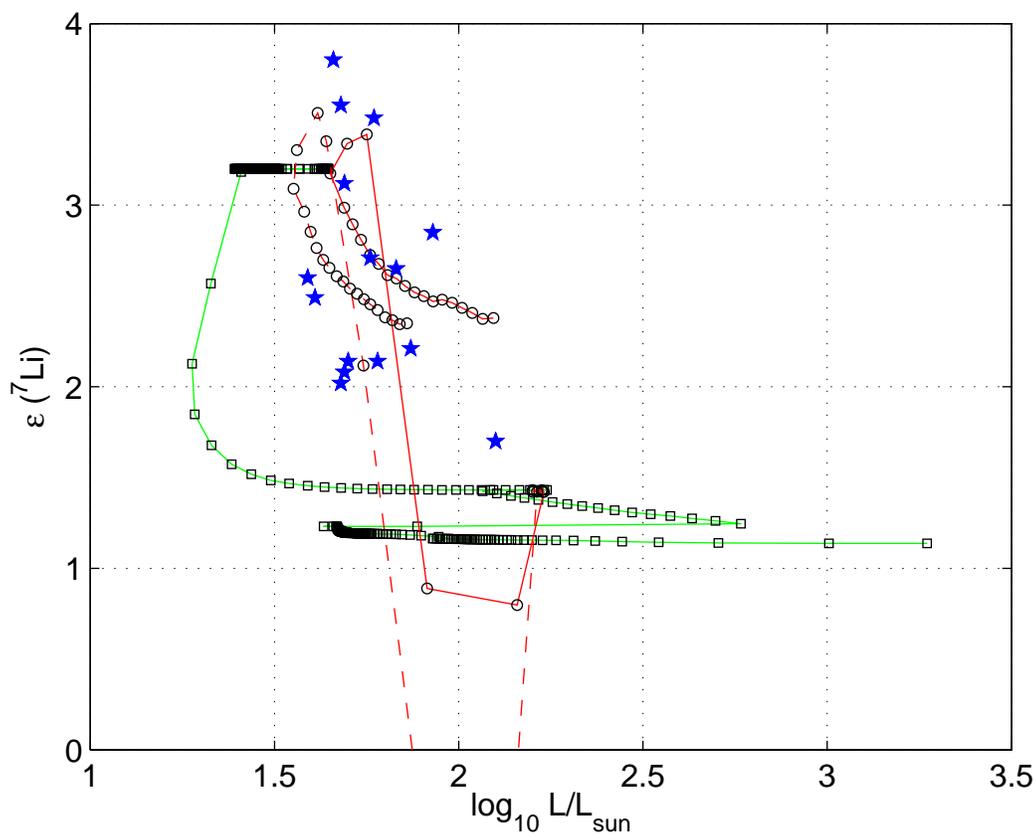}
\caption{The solid and dashed red curves show the changes of the surface Li abundance produced by enhanced extra mixing
         in the radiative zone of the $2.2\,M_\odot$ star above the bump luminosity. 
         It is assumed that $D_{\rm mix} = 2\times 10^{11}$ cm$^2$\,s$^{-1}$,
         while $\Gamma = 0.75$ and $\Gamma = 1$ in the first and second cases. The green curve represents
         the evolutionary variations of the Li abundance caused by the FDU and thermohaline convection with the fluid
         element's aspect ratio $a = 10$. The blue star symbols are the Li-rich K giants.
         }
\label{fig:f2}
\end{figure}

\begin{figure}
\epsffile [60 190 480 695] {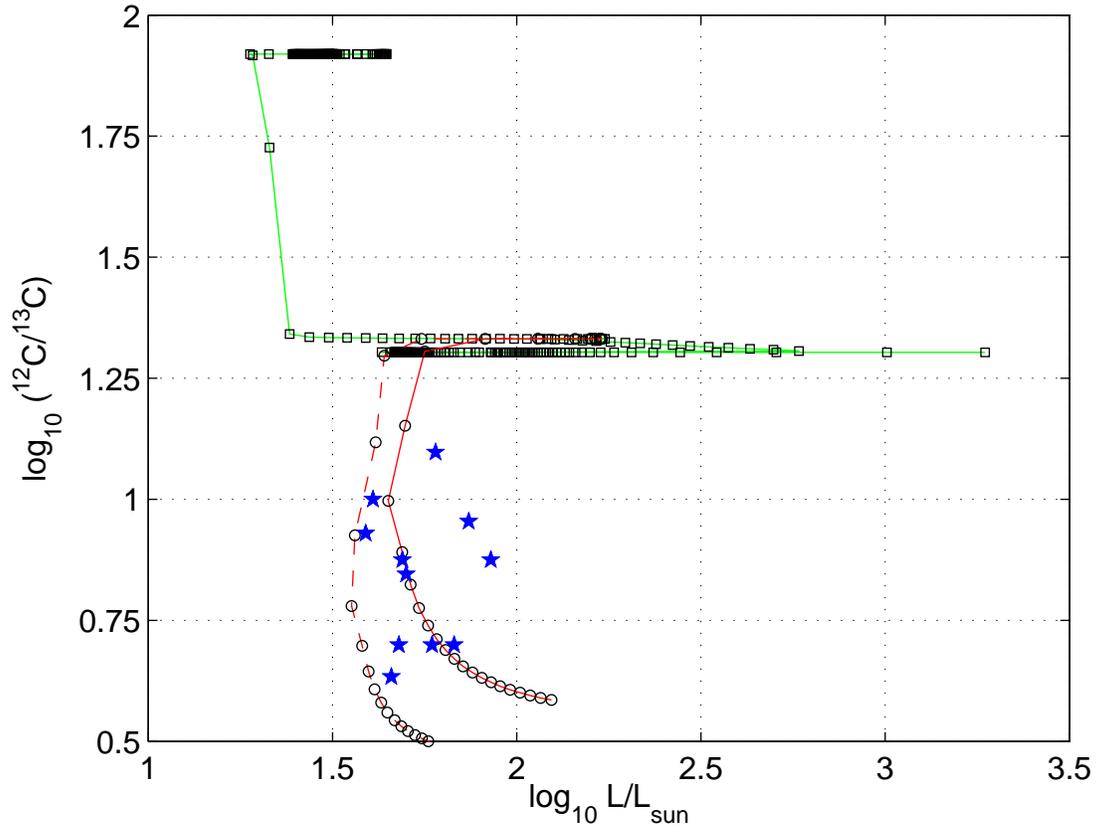}
\caption{The same as in Fig.~\ref{fig:f2}, but for the surface carbon isotopic ratio.
         }
\label{fig:f3}
\end{figure}

The parameter values $D_{\rm mix} = 2\times 10^{11}$\,cm$^2$\,s$^{-1}$, $r_{\rm mix} = 0.05\,R_\odot$, and $\Gamma = 0.75$ 
lead to both Li enrichment\footnote{For the Li abundance, we use the notation $\varepsilon(^7\mbox{Li})\equiv\log_{10}
[n(^7\mbox{Li})/n(\mbox{H})] + 12$, where $n$ is the number density of the corresponding nucleus.} 
and reduction of the $^{12}$C/$^{13}$C ratio consistent with those reported by KRL for the Li-rich 
K giants (solid red curves in Fig.~\ref{fig:f2} and Fig.~\ref{fig:f3}). Furthermore, an increase of
the efficiency of convective heat transport to $\Gamma = 1$ shifts the HRD location of the Li enrichment and, especially, 
that of the $^{12}$C/$^{13}$C sharp decline to the edges of their corresponding observational domains (dashed red curves 
in the same figures). Therefore, if our hypothesis is correct, $\Gamma$ should be close to 0.75 and not exceed 1 by much.

A similar value of $\Gamma$ is needed to explain the photometric peculiarity of red giants in tidally locked binaries,
including the RS CVn giant components (Fig.~\ref{fig:f1}). \cite{dea06} have predicted that some of these stars, those
that have already reached the bump luminosity and are now making extended zigzags towards lower luminosities as a result of
their fast rotation, should have carbon isotopic ratios smaller than the standard post-FDU value of
$^{12}$C/$^{13}\mbox{C}\approx 25$. Recently, two such stars have actually been found (\citealt{tea10,bea10});
the chromospherically active RS CVn-type stars $\lambda$\,And (HD\,222107) and 29 Draconis (HD\,160538) 
are located below the bump luminosity but have $^{12}$C/$^{13}\mbox{C}=14$ and 16, respectively. 

When comparing the Li-rich K giants with the RS CVn binaries, it is important to bear in mind that a hunt after the former
is biased towards finding the stars that have already reached the bump luminosity and are now experiencing enhanced extra mixing,
while a search for the latter selects both the pre- and post-bump luminosity stars. Therefore, it is not surprising that
the third RS CVn-type star, 33 Piscium, studied by the same group does not show signatures of extra mixing (\citealt{bea11}).
It would be interesting to measure carbon isotopic ratios in the following RS CVn stars: HD\,19754, HD\,182776,
HD\,202134, HD\,204128, and HD\,205249. They have relatively high (for red giants) projected rotational velocities 
($v\sin i > 7$\,km\,s$^{-1}$) and probably strongly increased atmospheric abundances of Na and Al (\citealt{mea04}), 
which might be a manifestation of enhanced extra mixing (\citealt{dvb03a}).

For comparison, green curves in Fig.~\ref{fig:f2} and Fig.~\ref{fig:f3} show the evolutionary changes of the Li abundance
and $^{12}$C/$^{13}$C ratio at the surface of the $2.2\,M_\odot$ star produced by the FDU (before the bump luminosity) and
thermohaline convection driven by $^3$He burning that reduces $\mu$ in the tail of the HBS
after the bump luminosity (\citealt{chz07}). The rate of the latter is approximated by
$D_\mu = 2\pi^2a^2[\nabla_\mu/(\nabla_{\rm rad}-\nabla_{\rm ad})]K$ (\citealt{d10a}) 
with the observationally constrained value of the aspect ratio, $a\equiv l/d=10$, of a fluid element,
where $l$ and $d$ are its length and diameter. Note that direct numerical simulations of the $^3$He-driven thermohaline convection
in a bump-luminosity red giant give an estimate of $a\approx 1$, which does not support the hypothesis that it is
the principal mechanism of RGB extra mixing (\citealt{d10a,dm11,tea11}).

\section{Discussion}
\label{sec:disc}

\begin{figure}
\epsffile [60 190 480 695] {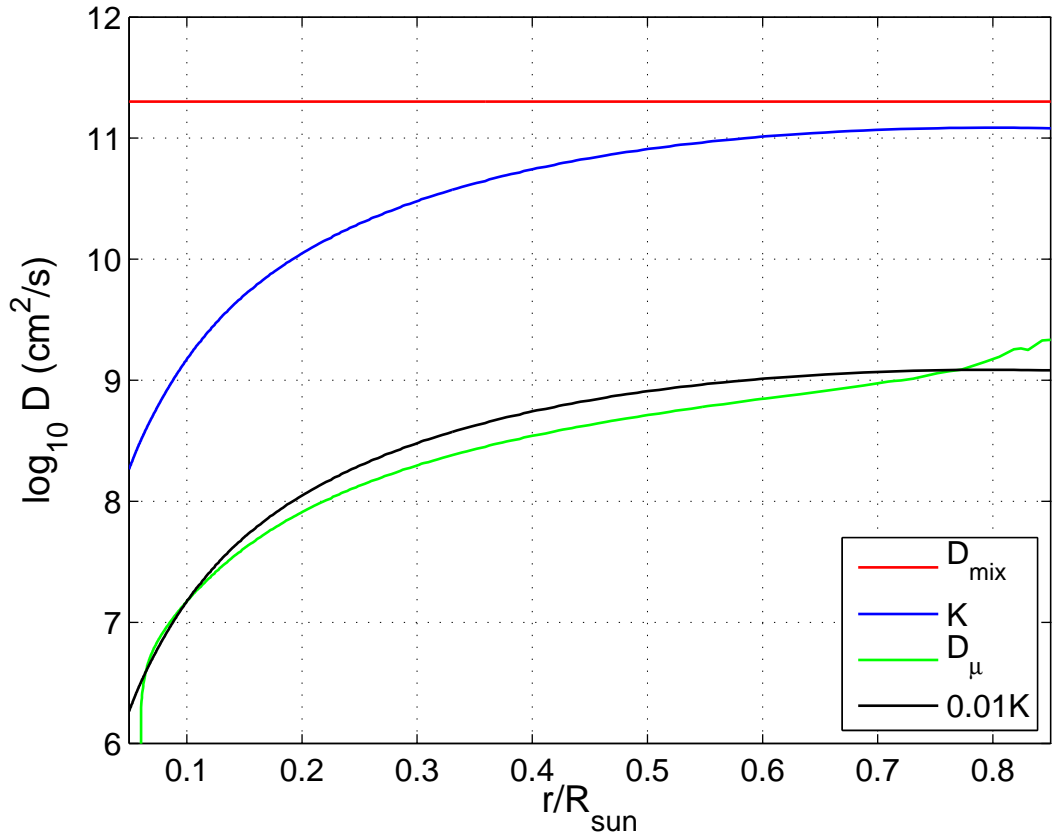}
\caption{The profiles of different diffusion coefficients in the radiative zone of the $2.2\,M_\odot$ star at the bump luminosity.
         }
\label{fig:f4}
\end{figure}

For the majority of low-mass red giants, that are experiencing extra mixing above the bump luminosity, the observed evolutionary
changes of their surface chemical composition are reproduced reasonably well either with the above-mentioned thermohaline
diffusion coefficient
$D_\mu$ that uses $a=10$, or with the diffusion coefficient $D_{\rm mix}\approx 0.01K$\,--\,$0.1K$ and appropriately 
chosen mixing depth, e.g. $r_{\rm mix} = 0.05\,R_\odot$ (\citealt{dp08a,dp08b}). For the radiative zone of our $2.2\,M_\odot$
bump-luminosity model, these diffusion coefficients are plotted in Fig.~\ref{fig:f4} 
(the green and black curves). Their corresponding efficiencies of heat transport are negligibly small. Therefore, in the majority of
cases, the (canonical) RGB extra mixing should not lead to noticeable changes in the evolution of these stars on the HRD.

On the contrary, the value of $D_{\rm mix} = 2\times 10^{11}$ cm$^2$\,s$^{-1}$, that
is required to understand the origin of Li-rich K giants in our models, results in $\Gamma = \gamma(D_{\rm mix}/K)\gg 1$,
at least in the inner half of the radiative zone
(compare the red and blue curves in Fig.~\ref{fig:f4}), therefore such (enhanced) extra mixing has to be accompanied by
an efficient heat transport. We have shown that the increase of $\Gamma$ to a value between 0.75 and 1
also helps to explain the photometric peculiarities intrinsic to both the majority of red giants in tidally locked binaries, 
including the RS CVn stars, and the Li-rich K giants studied by KRL. However, post-bump luminosity stars with $\Gamma\gg 1$
would make zigzags that are too lengthy, in conflict with observations. This disagreement can be resolved if, following
\cite{z92}, we assume that
the enhanced extra mixing is produced by highly anisotropic turbulence whose associated chemical and heat transport
in the horizontal direction strongly dominates over those in the vertical direction, $D_{\rm h}\gg D_{\rm mix}$.
\cite{dp08b} have demonstrated that, in this case, $\Gamma = D_{\rm mix}/[2a^2(K+D_{\rm h})]$ and
$f = 3\Gamma(D_{\rm mix}/K)/[1+\Gamma + 3\Gamma(D_{\rm mix}/K)]$, while the expression
$\nabla = (1-f)\nabla_{\rm rad} + f\nabla_{\rm ad}$ still holds. In the standard mixing length theory,
$\Gamma = 0.75$ and $\Gamma = 1$ correspond to $f= 0.66$ and $f=0.75$, respectively. For the anisotropic turbulent mixing,
we obtain $\Gamma \ll 1$, assuming that $D_{\rm h}\gg D_{\rm mix}\gg K$ and $a\geq 1$. At the same time, we can still
get the observationally supported value of $f\approx 0.75$, provided that $\Gamma(D_{\rm mix}/K)\approx 1$, or
$2(D_{\rm h}/K)\approx (D_{\rm mix}/K)^2$. This means that, for $D_{\rm mix}\approx 10^2K$ (Fig.~\ref{fig:f4}),
we need $D_{\rm h}\approx 5\times 10^3D_{\rm mix}$ to keep $f$ close to 0.75. Interestingly, \cite{d10b}
has used a similar ratio of $D_{\rm h}$ to $D_{\rm mix}$ for the rotation-driven turbulent diffusion to make his model of 
magnetic braking of solar rotation consistent with observational data. Moreover, the same order of magnitude
ratios $D_{\rm h}/D_{\rm mix}$ were obtained for radiative zones of bump luminosity red giant models by \cite{pea06} 
who took into account self-consistently the transport of angular momentum by rotation-driven meridional 
circulation and shear turbulence.

Given that $\nabla_{\rm rad}\approx 0.2$ and $\nabla_{\rm ad}\approx 0.4$
in the radiative zone of a bump-luminosity red giant,
the maximum deviation of $\nabla$ from $\nabla_{\rm rad}$ that can be achieved at the break-up rotation corresponds to
$f\approx 2\varepsilon_{\rm crit}=0.48$, which is too small to reproduce the extended zigzags presumably made
by the Li-rich K giants and RS CVn giant components. However, this conclusion is based only on the analysis of the modification
of the temperature gradient by the rotational distortion. It ignores the fact that effects of rotation are also
incorporated in other equations of stellar structure (\citealt{dvb03b}). With all the effects taken into account,
\cite{dea06} were actually able to construct the evolutionary track of a $1.7\,M_\odot$ star whose post-bump luminosity
zigzag reached the HRD domain occupied by the RS CVn binaries. 

Comparing the green and red curves in Fig.~\ref{fig:f4}, we infer that the $^3$He-driven thermohaline
convection could be responsible for the observed Li enrichment only if the aspect ratio of its fluid elements were as large as
$a\approx 300$. At present, we consider this highly improbable. The more likely interpretation from our point of view is
that the Li enrichment in the stars studied by KRL and their location on the lower RGB below the bump luminosity
are both caused by their fast internal rotation and its associated turbulent mixing and heat transport. This hypothesis is 
supported by the fact that surface rotation velocities of field stars with $M\ga 1.6\,M_\odot$ remain constant as they evolve
on and away from the main sequence (\citealt{ws97}). It is likely that
only the most rapidly rotating stars, of which a handful have $v\sin i > 200$ km\,s$^{-1}$,
become Li-rich and make extended zigzags after the bump luminosity.
Our hypothesis does not exclude the possibility of Li enrichment above the bump luminosity in a red giant that has been
spun up as a result of its engulfing an orbiting giant planet (\citealt{dw00}).

The only possible scenario in which a substantial mixing of H-rich material occurs during the He-core flash is
the ``hydrogen injection flash'' (\citealt{mea11}). The He ignition starts off-center, which leads to the formation of
a He convective shell. For this scenario to work, convection from the He-burning shell must penetrate into the HBS.
In the MESA stellar evolution code, such penetration, also known as convective overshooting, is modeled by
the exponentially decaying diffusion coefficient $D_{\rm OV} = D_0\exp[-2|r-r_0|/(fH_P)]$,
where $H_P$ is the pressure scale height, and $D_0$ is a convective diffusion coefficient estimated using
a mixing-length theory at the radius $r_0$ near the convective boundary (\citealt{hea97}).
Our computations of the He-core flash in the $1.7\,M_\odot$
star show that H can be injected into the He convective shell only when $f\ga 0.15$. This exceeds the 
observationally constrained value of $f$ by nearly an order of magnitude (\citealt{nea10}, and references therein).
Besides, given the discussed similarities between the Li-rich K giants studied by KRL and the RS CVn giant components
and the fact that the latter stars are definitely on the lower RGB, because otherwise
they would have undergone a common-envelope event with their close binary companions, we believe that these stars
are in the same evolutionary phase. It should also be noted that the inclusion of convective overshooting in main-sequence 
stars would shift the maximum initial mass of the solar-composition stars that experience the He-core flash 
from $M\approx 2.2\,M_\odot$ to $M\approx 1.8\,M_\odot$.

Our hypothesis could eventually be tested by asteroseismology
because high-precision photometry capable of measuring g-mode oscillations in red giants allows ``to distinguish unambiguously 
between hydrogen-shell-burning stars (period spacing mostly $\sim$\,50 seconds) and those that are also burning helium 
(period spacing $\sim$\,100 to 300 seconds)'' (\citealt{beddea11}).

\acknowledgements
I am thankful to Falk Herwig and Don VandenBerg for useful comments and for supporting my work through their Discovery Grants
from Natural Sciences and Engineering Research Council of Canada.


\begin{thebibliography}{}

\bibitem[Barisevi\u{c}ius et al.(2010)]{bea10}
Barisevi\u{c}ius, G., Tautvai\u{s}ien\.{e}, G., Berdyugina, S., Chornyi, Y., \& Ilyin, I.~2010, BaltA, 19, 157 

\bibitem[Barisevi\u{c}ius et al.(2011)]{bea11}
Barisevi\u{c}ius, G., Tautvai\u{s}ien\.{e}, G., Berdyugina, S., Chornyi, Y., \& Ilyin, I.~2010, BaltA, 20, 53  

\bibitem[Bedding et al.(2011)]{beddea11}
Bedding, S., Mosser, B., Huber, D., et al.~2011, Nature, 471, 608

\bibitem[Cameron \& Fowler(1971)]{cf71}
Cameron, A.~G.~W., \& Fowler, W.~A.~1971, ApJ, 164, 111

\bibitem[Charbonnel \& Balachandran(2000)]{chb00}
Charbonnel, C., \& Balachandran, S.~C.~2000, A\&A, 359, 563

\bibitem[Charbonnel \& Zahn(2007)]{chz07}
Charbonnel, C., \& Zahn, J.-P.~2007, A\&A, 467, L15

\bibitem[Charbonnel \& Lagarde(2010)]{chl10}
Charbonnel, C., \& Lagarde, N.~2010, A\&A, 522, A10

\bibitem[Denissenkov(2010a)]{d10a}
Denissenkov, P.~A.~2010a, ApJ, 723, 563

\bibitem[Denissenkov(2010b)]{d10b}
Denissenkov, P.~A.~2010b, ApJ, 719, 28 

\bibitem[Denissenkov, Chaboyer, \& Li (2006)]{dea06}
Denissenkov, P.~A., Chaboyer, B., \& Li, K.~2006, ApJ, 641, 1087

\bibitem[Denissenkov, \& Weiss (2000)]{dw00}
Denissenkov, P.~A., \& Weiss, A.~2000, A\&A, 358, L49 

\bibitem[Denissenkov, \& VandenBerg (2003a)]{dvb03a}
Denissenkov, P.~A., \& VandenBerg, D.~A.~2003a, ApJ, 593, 509 

\bibitem[Denissenkov, \& VandenBerg (2003b)]{dvb03b}
Denissenkov, P.~A., \& VandenBerg, D.~A.~2003b, ApJ, 598, 1246

\bibitem[Denissenkov, \& Pinsonneault(2008a)]{dp08a}
Denissenkov, P.~A., \& Pinsonneault, M.~2008a, ApJ, 679, 1541

\bibitem[Denissenkov, \& Pinsonneault(2008b)]{dp08b}
Denissenkov, P.~A., \& Pinsonneault, M.~2008b, ApJ, 684, 626

\bibitem[Denissenkov, \& Merryfield(2011)]{dm11}
Denissenkov, P.~A., \& Merryfiled, W.~J.~2011, ApJ, 727, L8 

\bibitem[Fekel et al.(2002)]{fea02}
Fekel, F.~C., Henry, G.~W., Eaton, J.~A., Sperauskas, J., \& Hall, D.~S.~2002, AJ, 124, 1064

\bibitem[Fekel \& Henry(2005)]{fh05}
Fekel, F.~C., \& Henry, G.~W.~2005, AJ, 129, 1669

\bibitem[Gratton et al.(2000)]{grea00}
Gratton, R.~G., Sneden, C., Carretta, E., \& Bragaglia, A.~2000, A\&A, 354, 169

\bibitem[Herwig et al.(1997)]{hea97}
Herwig, F., Bl\"{o}cker, T., Sch\"{o}nberner, D., \& El Eid, M.~1997, A\&A, 324, L81

\bibitem[Kumar, Reddy, \& Lambert (2011)]{kea11}
Kumar, Y.~B., Reddy, B.~E., \& Lambert, D.~L.~2011, ApJ, 730, L12 (KRL)

\bibitem[Maeder(1995)]{m95}
Maeder, A. 1995, A\&A, 299, 84   

\bibitem[Moc\'{a}k, Siess, \& M\"{u}ller (2011)]{mea11}
Moc\'{a}k, M., Siess, L., \& M\"{u}ller, E.~2011, A\&A, 533, 53   

\bibitem[Morel et al.(2004)]{mea04}
Morel, T., Micela, G., Favata, F., \& Katz, D.~2004, A\&A, 426, 1007

\bibitem[Noels et al.(2010)]{nea10}
Noels, A., Montalban, J., Miglio, A., Godart, M., \& Ventura, P.~2010, Ap\&SS, 328, 227  

\bibitem[Palacios et al.(2006)]{pea06}
Palacios, A., Charbonnel, C., Talon, S., \& Siess, L.~2006, A\&A, 453, 261

\bibitem[Paxton et al.(2011)]{pea11}
Paxton, B., Bildsten, L., Dotter, A., Herwig, F., Lessafre, P., \& Timmes, F.~2011, ApJS, 192, 3

\bibitem[Sackmann, \& Boothroyd (1999)]{sb99}
Sackmann, I.-J., \& Boothroyd, A.~I.~1999, ApJ, 510, 217 

\bibitem[Spruit(1999)]{s99}
Spruit, H.~C.~1999, A\&A, 349, 189

\bibitem[Tautvai\u{s}ien\.{e} et al.(2010)]{tea10}
Tautvai\u{s}ien\.{e}, G., Barisevi\u{c}ius, G., Berdyugina, S., Chornyi, Y., \& Ilyin, I.~2010, BaltA, 19, 95  

\bibitem[Traxler, Garaud, \& Stellmach (2011)]{tea11}
Traxler, A., Garaud, T., \& Stellmach, S.~2011, ApJ, 728, L29 

\bibitem[Wolff, \& Simon (1997)]{ws97}
Wolff, S.-C., \& Simon, T.~1997, PASP, 109, 759 

\bibitem[Zahn(1992)]{z92}
Zahn, J.-P.~1992, A\&A, 256, 115

\end{thebibliography}
\end{document}